\documentclass[12pt,a4paper]{article}
\usepackage[utf8x]{inputenc}

\usepackage{amsmath}
\usepackage{amsfonts}
\usepackage{amssymb}

\usepackage{graphicx}

\usepackage[perpage]{footmisc}

\usepackage[bookmarks=true,colorlinks=true,citecolor=blue,linkcolor=red,urlcolor=magenta]{hyperref}

\usepackage{multirow}

\usepackage[table]{xcolor}

\usepackage{geometry} 
\begin{document}

\author{V.D.~Rusov\footnote{Corresponding author: Vitaliy D. Rusov, E-mail: siiis@te.net.ua}, V.A.~Tarasov, S.I.~Kosenko, S.A.~Chernegenko}

\title{The resonance absorption probability function for neutron and multiplicative integral}

\date{}

\maketitle

\begin{center}
    \textit{Department of Theoretical and Experimental Nuclear Physics, \\ Odessa  National Polytechnic University, Odessa, Ukraine}
\end{center}

\abstract{The analytical approximations for the moderating neutrons flux density like Fermi spectra, widely used in reactor physics, involve the probability function for moderating neutron to avoid the resonant absorption obtained using some restrictive assumptions regarding the acceptable resonances width. By means of multiplicative integral (Volterra integral) theory for a commutative algebra an analytical expression for the probability function is obtained rigorously without any restrictive assumptions.}


\section{Introduction}
\label{sec1}

The compound-nuclear reactions may be divided into resonant and non resonant. It is known that the nucleus energy may possess only a discrete set of values corresponding to its energy levels. However the idea of the levels with strictly fixed energy is true only for the ground levels (non-excited levels) of the stable nuclei. All the other (excited) levels do not possess the fixed energy – they are more or less smeared over energy. A width $\Gamma_i$ ($i$ is the level index) of such smearing can be estimated using the uncertainty relation for time and energy. According to this estimate, $\Delta E_i = \Gamma_i = \hbar / \tau_i$, where $\tau_i$ is the lifetime for current level. The nucleus may be excited through the interaction with some projectile particle only if this excitation energy induced by interaction corresponds to the energy levels spacing. Therefore a compound nucleus may form only in the case that the projectile particle energy is such that the excitation energy of the compound nucleus induced by this interaction fits into the corresponding level's interval of uncertainty $\Gamma_i$ (if the nucleus had been in its ground state prior to interaction). If the width of compound nucleus energy levels is less than the distance between them and the projectile particles possess some fixed energy, the reaction may take place only trough some $i^{th}$ single level. The reaction cross-section dependence on projectile particle energy would display a resonant behavior. Correspondingly, this kind of reactions is called resonant. If the levels are located so densely that the distances between them is less than their width, they merge together and the reaction may take place for any projectile particle energy. This kind of reactions is called non-resonant. The form of resonances is known to be described by the Breit–Wigner function.

\begin{figure}
  \begin{center}
    \includegraphics[width=10cm]{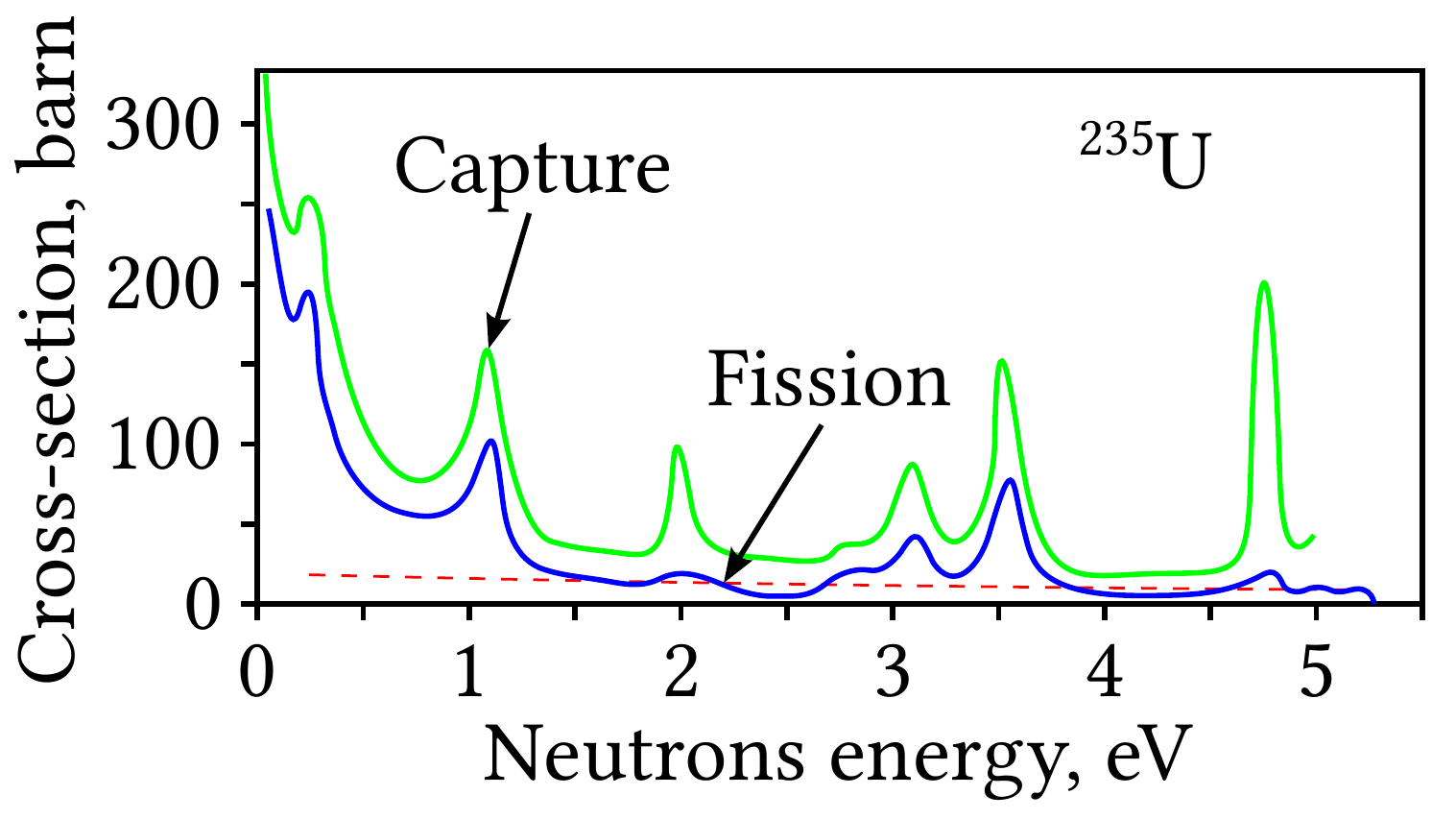}
  \end{center}
\caption{$^{235}U$ capture and fission cross-sections dependence on neutrons energy. Dashed line represents the neutrons scattering cross-section dependence.}
\label{fig1}
\end{figure}

As is known, the resonance region of the reactor neutron nuclear reactions of fissile nuclides is located, depending on neutrons energy, in the 0.5$\div$1000~eV range (e.g. \cite{ref3} and Fig.~\ref{fig1}).

Determining the neutrons spectrum for a moderating medium with resonant absorption remains an important task for numerous problems of nuclear reactor physics. Numerical methods of integro-differential equation solution for the moderating neutrons flux density in the medium with resonant absorption are used as well as the widespread analytical approximations for the flux density like Fermi spectra. Fermi spectra involve the probability function for moderating neutron to avoid the resonant absorption. The specific expression for this probability function is usually obtained using some restrictive assumptions regarding the acceptable resonances width. The narrow, broad (infinite mass approximation) and intermediate resonance approximations are used for that purpose. Therefore depending on the properties of nuclei composing the moderating medium with absorption, one of the above approximations is used.

We show in the present paper that the probability function for a moderating neutron to avoid the resonant absorption (let us note that the neutron absorption is the opposite event, and therefore the absorption probability function is equal to one minus the probability function to avoid the absorption) may be written down in a form of multiplicative integral (Volterra integral) \cite{ref1}. The analytical expression for this function was rigorously obtained using the multiplicative integral theory for a commutative algebra \cite{ref2} without any restricting assumptions.

\section{Probability function for avoiding the resonant capture with narrow resonance approximation (standard approach)}
\label{sec2}

Fermi spectrum for a moderating medium with resonant absorption, according to \cite{ref3,ref4}, has the form:

\begin{equation}
\Phi_{\Phi} (E) = \dfrac{S}{\overline{\xi} \Sigma_t E} \cdot \varphi (E),
\label{eq1}
\end{equation}

\noindent where $\varphi (E)$ is the probability function for a moderating neutron to avoid the resonant absorption; $S$ is the total  volume neutron generation rate; $\overline{\xi} = \sum \limits_i \left( \xi_i \Sigma_S^i \right) / \Sigma_S$, $\xi_i$ is the average logarithmic decrement of  energy loss; $\Sigma_S^i$ is the macroscopic scattering cross-section and $\Sigma_a^i$ is the macroscopic absorption cross-section of the $i^{th}$ nuclide; $\Sigma_t = \sum \limits_i \Sigma_S^i + \Sigma_a^i$ is the total macroscopic cross-section of the fissile material, $\Sigma_S = \sum \limits_i \Sigma_S^i$ is the total macroscopic scattering cross-section of the fissile material.

\begin{figure}
  \begin{center}
    \includegraphics[width=10cm]{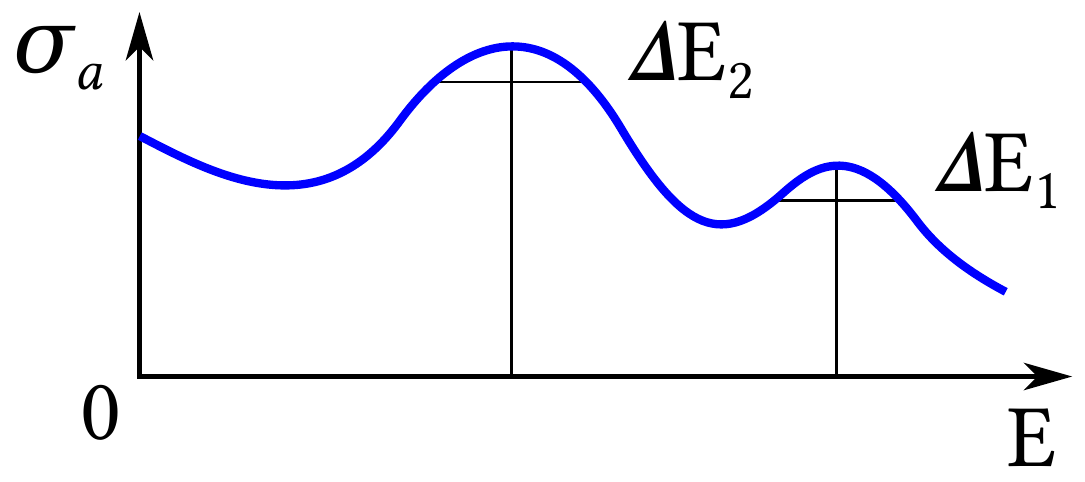}
  \end{center}
\caption{Two distant resonances $E_1$ and $E_2$ ($\Delta E_1$, $\Delta E_2$ are the widths of the first and second resonances respectively).}
\label{fig2}
\end{figure}

A following standard approach is used in reactor physics \cite{ref3,ref4} in order to obtain the probability function $\varphi (E)$. According to the standard slowing-down theory and Fig.~\ref{fig2}, which represents the resonance region for neutron absorption cross-section in a simplified visual manner, the probability for neutron to avoid the absorption within the first resonance region may be written in the form:

\begin{equation}
\varphi_1 = 1 - \dfrac{\Sigma_a}{\overline{\xi} \left( \Sigma_S + \Sigma_a \right)} \dfrac{\Delta E_1}{E_1},
\label{eq2}
\end{equation}

\noindent where $\Sigma_a$ is the macroscopic absorption cross-section; $E_1$ is the first resonance energy and $\Delta E_1$ is its width (see Fig.~\ref{fig2}).

The probability for neutron to avoid the absorption within the second resonance region with energy $E_2$ and width $\Delta E_2$ has the following form:

\begin{equation}
\varphi_2 = \left[ 1 - \dfrac{\Sigma_a}{\overline{\xi} \left( \Sigma_S + \Sigma_a \right)} \dfrac{\Delta E_1}{E_1} \right] \times \left[ 1 - \dfrac{\Sigma_a}{\overline{\xi} \left( \Sigma_S + \Sigma_a \right)} \dfrac{\Delta E_2}{E_2} \right].
\label{eq3}
\end{equation}

Thereby by analogy we have the probability function for a moderating neutron to avoid the resonant absorption:

\begin{equation}
\varphi (E) = \prod \limits_{i=1}^N \varphi_i = \prod \limits_{i=1}^N \varphi (E_i) = \prod \limits_{i=1}^N \left(  1 - \dfrac{\Sigma_a}{\overline{\xi} \left( \Sigma_S + \Sigma_a \right)} \dfrac{\Delta E_i}{E_i} \right),
\label{eq4}
\end{equation}

\noindent where $N$ is a number of resonances.

In order to pass from multiplication to summation, we take a logarithm of the function $\varphi (E)$ defined in (\ref{eq4}):

\begin{equation}
\ln {\varphi (E)} = \prod \limits_{i=1}^N \ln \left(  1 - \dfrac{\Sigma_a}{\overline{\xi} \left( \Sigma_S + \Sigma_a \right)} \dfrac{\Delta E_i}{E_i} \right) = \prod \limits_{i=1}^N \ln \left(  1 - x_i \right),
\label{eq5}
\end{equation}

\noindent where $x_i = \dfrac{\Sigma_a}{\overline{\xi} \left( \Sigma_S + \Sigma_a \right)} \dfrac{\Delta E_i}{E_i}$.

Next every multiplicand in expression (\ref{eq5}) is expanded into series in terms of a small quantity $x_i$ with an assumption that $\dfrac{\Delta E_i}{E_i} \ll 1$. Limiting to the first expansion term, we obtain the following expression:

\begin{equation}
\ln \varphi (E) \approx - \sum \limits_{i=1}^N \dfrac{\Sigma_a}{\overline{\xi} \left( \Sigma_S + \Sigma_a \right)} \dfrac{\Delta E_i}{E_i}.
\label{eq6}
\end{equation}

Let us divide the resonance region into $m$ energy intervals of width $\Delta E_j$ and assume $\Sigma_a = 0$ in the intervals between the resonances. Thus for (\ref{eq6}) we obtain:

\begin{equation}
\ln \varphi (E) \approx - \sum \limits_{j=1}^m \dfrac{\Sigma_a}{\overline{\xi} \left( \Sigma_S + \Sigma_a \right)} \dfrac{\Delta E_j}{E_j}.
\label{eq7}
\end{equation}

Passing from the logarithm to an exponent and making $m$ approach the infinity, for entire resonance region from $E_0$ to $E_f$ we have:

\begin{equation}
\varphi (E) \approx \lim \limits_{m \to \infty} \exp \left( - \sum \limits_{j=1}^m \dfrac{\Sigma_a}{\overline{\xi} \left( \Sigma_S + \Sigma_a \right)} \dfrac{\Delta E_j}{E_j} \right) = \exp \left( - \int \limits_{E_0}^{E_f} \dfrac{\Sigma_a}{\overline{\xi} \left( \Sigma_S + \Sigma_a \right)} \dfrac{dE}{E} \right).
\label{eq8}
\end{equation}

The Fermi spectrum (\ref{eq1}) with the probability function $\varphi (E)$ in exponential form (\ref{eq8}) is called a narrow resonance approximation.

\section{Multiplicative integral and its application to a commutative algebra $A$}
\label{sec3}

Basic structures related to a multiplicative integral occur in different areas of mathematics, mechanics, physics and -- as it follows from the sections \ref{sec1} and \ref{sec2} above -- in the nuclear reactions and nuclear reactor physics as well. However one cannot say that  the multiplicative integral theory is currently "particularly popular" among mathematicians and physicists. Let us define multiplicative integral according to \cite{ref1}. Let $A$ be an arbitrary associative topological algebra with an identity $E$, and $f(t)$ is a function of real variable $t$ which takes on the values in the algebra $A$, $a,b \in \mathbf{R}, a \leqslant b, [a,b]=\lbrace t \in \mathbf{R}, a \leqslant t \leqslant b \rbrace$, $T$ is the partition of an interval $[a,b]$ with the points $t_0 = a, t_1, t_2, \ldots, t_{n-1}, t_n = b$, $t_i \leqslant t_{i+1}, i = 0,1,2, \ldots, n-1$; $l(T) = \max (t_{i+1} - t_i)$, $\Delta t_i = t_{i+1}-t_i$.

Let us consider the following product:

\begin{equation}
\prod \left( f, T \right) = \left( E + 	f(t_0) \Delta t_0 \right) \times \left( E + 	f(t_1) \Delta t_1 \right) \times \ldots \times \left( E + 	f(t_n) \Delta t_n \right).
\label{eq9}
\end{equation}

If for any variation T such that $l(T) \to 0$ the product $\prod \left( f, T \right)$ converges to some limit, this limit is called a \textit{multiplicative integral} of the function $f(t)$ over the interval $[a,b]$ and is denoted by

\begin{equation}
\int \limits_a^{^{\cup} b} E + f(t) dt.
\label{eq10}
\end{equation}

By analogy, for $f(t)$ and $T$ one can derive the following:

\begin{equation}
^{\cap} \prod \left( f, T \right) = \left( E + 	f(t_n) \Delta t_n \right) \times \left( E + 	f(t_{n-1}) \Delta t_{n-1} \right) \times \ldots \times \left( E + 	f(t_0) \Delta t_0 \right).
\label{eq11}
\end{equation}

If for any variation T such that $l(T) \to 0$ the product $^{\cap}\prod \left( f, T \right)$ converges to some limit, this limit gives another type of the multiplicative integral for the function $f(t)$ over the interval $[a,b]$ and is denoted by

\begin{equation}
\int \limits_a^{^{\cap} b} E + f(t) dt.
\label{eq12}
\end{equation}

The multiplicative integrals (\ref{eq10}) and (\ref{eq11}) are called \textit{direct} and \textit{backward} respectively.

Next let us consider a particular case of multiplicative integral application. Let the algebra $A$ be commutative. Then, according to \cite{ref1,ref2},

\begin{align}
& \int \limits_a^{^{\cap} b} E + f(t) dt = \lim \limits_{l(T) \to 0}  \left( E + 	f(t_0) \Delta t_0 \right) \times \left( E + 	f(t_1) \Delta t_1 \right) \times \ldots \times \left( E + 	f(t_n) \Delta t_n \right) = \nonumber \\
 & = \lim \limits_{l(T) \to 0} \left[ E + \sum f(t_i) \Delta t_i + \sum \limits_{i<j} f(t_i) f(t_j) \Delta t_i \Delta t_j +  \sum \limits_{i<j<k} f(t_i) f(t_j) f(t_k) \Delta t_i \Delta t_j \Delta t_k + \ldots  \right] = \nonumber \\
 & = \lim \limits_{l(T) \to 0} \left[ \sigma_0 + \sigma_1 + \sigma_2 + \ldots + \sigma_n \right],
\label{eq13}
\end{align}

\noindent where $\sigma_0, \sigma_1, \sigma_2, \ldots, \sigma_n$ are the elementary symmetric polynomials in variables $f(t_0) \Delta t_0$, $f(t_1) \Delta t_1$, $f(t_2) \Delta t_2$, $\ldots$,  $f(t_n) \Delta t_n$. Let $\rho_0$, $\rho_1$, $\rho_2$, $\ldots$, $\rho_n$ be symmetric Newton's polynomials in the same variables:

\begin{equation}
\rho_k = \sum \limits_i \left[ f(t_i) \Delta t_i \right] ^k, ~~ k=1,2,\ldots,n.
\label{eq14}
\end{equation}

It is easy to see that

\begin{equation}
\lim \limits_{l(T) \to 0} \rho_k = 
\begin{cases}
0, & k > 0, k \neq 1 \\
\int \limits_a^b f(t) dt, & k=1,
\end{cases}
\label{eq15}
\end{equation}

\noindent where $\int \limits_a^b f(t) dt$ is the Riemann integral.

Expanding the polynomials $\sigma_k$ in terms of $\rho_l$ ($k,l = 1,2,\ldots,n$) and applying the equality (\ref{eq15}), we calculate

\begin{equation}
\lim \limits_{l(T) \to 0} \sigma_k.
\label{eq16}
\end{equation}

It is known that 

\begin{equation}
\rho_m - \rho_{m-1}\sigma_1 + \rho_{m-2}\sigma_2 + \ldots + \rho_1\sigma_{m-1}(-1)^{m-1} + (-1)^m \sigma_m = 0, ~~~m=1,2,\ldots,n,
\label{eq17}
\end{equation}

\noindent whence it follows that

\begin{align}
\lim \sigma_1 & = \lim \rho_1 = \int \limits_a^b f(t) dt, \nonumber \\
\lim \sigma_2 & = \frac{1}{2!} \left( \int \limits_a^b f(t) dt \right)^2, \nonumber \\
& \ldots, \nonumber \\
\lim \sigma_n &=  \frac{1}{n!} \left( \int \limits_a^b f(t) dt \right)^n
\label{eq18}
\end{align}

\noindent and as a result (taking into account that $e^x = 1 + \frac{x}{1!} + \frac{x^2}{2!} + \frac{x^3}{3!} + \ldots + \frac{x^n}{n!} + \ldots$ and that $n \to \infty$ when $l(T) \to 0$) we obtain

\begin{equation}
\int \limits_a^{^{\cup}b} E + f(t) dt = \exp \left( \int \limits_a^b f(t) dt \right).
\label{eq19}
\end{equation}

Due to Eq.(\ref{eq13}) symmetry, for the backward multiplicative integral we, obviously, obtain the following equality:

\begin{equation}
\int \limits_a^{^{\cap}b} E + f(t) dt = \exp \left( \int \limits_a^b f(t) dt \right).
\label{eq20}
\end{equation}

\section{Multiplicative integral and its application to the derivation of a probability function for a moderating neutron to avoid the resonant absorption}
\label{sec4}

According to the stated above, it is easy to establish the correspondence between the theory of a probability function for a moderating neutron to avoid the resonant absorption which is a real scalar function (and therefore a particular case of commutative algebra $A$) and a multiplicative integral. Indeed, according to the section \ref{sec3}, the identity $E$ of this commutative algebra $A$ is the real unit  1, and $f(E) = - \dfrac{\Sigma_a}{\overline{\xi} \left( \Sigma_S + \Sigma_a \right) E}$ is a scalar function of a real argument $E$ which takes on the values in the algebra $A$, $E_0, E_f \in \mathbf{R}$, $E_0 < E_f$, $[E_0,E_f] = \lbrace E \in \mathbf{R}, E_0 \leqslant E \leqslant E_f \rbrace$, $T$ is the partition of an interval $[E_0,E_f]$ with the points $t_0 = E_0, t_1, t_2, \ldots, t_{n-1}, t_n = E_f$, $t_i \leqslant t_{i+1}, i = 0,1,2, \ldots, n-1$; $l(T) = \max (t_{i+1} - t_i)$, $\Delta t_i = t_{i+1}-t_i$.

Then for the probability function to avoid the resonant absorption, according to (\ref{eq4}) with the partition $T$ of an interval $[E_0,E_f]$ we obtain:

\begin{equation}
\varphi (E) = \lim \limits_{n \to \infty} \prod \limits_{i=1}^n \left( 1 - \dfrac{\Sigma_a}{\overline{\xi} \left( \Sigma_S + \Sigma_a \right) E_i} \Delta E_i \right) = \lim \limits_{l(T) \to \infty} \prod \limits_{i=1}^{n} \left( 1 + f(E_i) \Delta E_i \right) = \int \limits_{E_0}^{^{\cap}E_f} E + f(t) dt,
\label{eq21}
\end{equation}

\noindent where $f(E) = - \dfrac{\Sigma_a}{\overline{\xi} \left( \Sigma_S + \Sigma_a \right) E}$.

Consequently, taking into account (\ref{eq20}), we derive the final expression:

\begin{equation}
\varphi (E) = \exp \left( - \int \limits_E^{E_f} \dfrac{\Sigma_a}{\overline{\xi} \left( \Sigma_S + \Sigma_a \right) E} dE \right).
\label{eq22}
\end{equation}

\section{Conclusion}
\label{sec5}

In the present paper we showed that the probability function for a moderating neutron to avoid resonant absorption may be written down in a form of a multiplicative integral. Using the results of the multiplicative integral theory application for a commutative algebra the analytical expression for the probability function for a moderating neutron to avoid the resonant absorption was rigorously obtained. The expression for Fermi spectrum incorporating the obtained probability function turned out to coincide with the expression for Fermi spectrum known as the narrow resonance approximation. However, the  former one is an exact expression and therefore does not have any restrictions.

We should also note here that since the narrow resonance approximation is used for another types of nuclear reactions, the obtained results may be generalized for a broader range of nuclear physics problems.


\end{document}